\documentclass{PoS}

\usepackage{url}
\usepackage{graphicx}
\newcommand{\mc}[3]{\multicolumn{#1}{#2}{#3}}
\usepackage{xhfill}% http://ctan.org/pkg/xhfill

\title{{\em Robopol}: Optical polarisation monitoring of blazars}

\ShortTitle{Optical polarisation monitoring of blazars}

\author{\speaker{Emmanouil Angelakis}\thanks{Presented on behalf of the entire {\em RoboPol} collaboration
    (see note at the end of the manuscript)}\\
         Max-Planck-Institut f\"ur Radioastronomie, Auf dem H\"ugel 69, Bonn 53121, Germany \\
        E-mail: \email{eangelakis@mpifr.de}}

\author{Dmitry Blinov\\
        Department of Physics and Institute for Plasma Physics, University of Crete, 71003, Heraklion, Greece\\
        Foundation for Research and Technology - Hellas, IESL, Voutes, 71110 Heraklion, Greece\\
        Astronomical Institute, St. Petersburg State University,Universitetsky pr. 28, Petrodvoretz, 198504 St. Petersburg, Russia\\
        E-mail: \email{blinov@physics.uoc.gr}}

\author{Markus B\"ottcher\\
  North-West University, Potchefstroom Campus, Private Bag X6001, Potchefstroom 2520, South Africa\\
        E-mail: \email{Markus.Bottcher@nwu.ac.za}}

\author{Talvikki Hovatta\\
        Aalto University Mets\"ahovi Radio Observatory, Mets\"ahovintie 114, 02540 Kylm\"al\"a, Finland\\
        E-mail: \email{talvikki.hovatta@aalto.fi}}

\author{Sebastian Kiehlmann\\
        Aalto University Mets\"ahovi Radio Observatory, Mets\"ahovintie 114, 02540 Kylm\"al\"a, Finland\\
        E-mail: \email{Sebastian.kiehlmann@aalto.fi}}

\author{Ioannis Myserlis\\
         Max-Planck-Institut f\"ur Radioastronomie, Auf dem H\"ugel 69, Bonn 53121, Germany \\
        E-mail: \email{imyserlis@mpifr-bonn.mpg.de}}

\author{Vassiliki Pavlidou\\
        Department of Physics and Institute for Plasma Physics, University of Crete, 71003, Heraklion, Greece\\
        Foundation for Research and Technology - Hellas, IESL, Voutes, 71110 Heraklion, Greece\\
        E-mail: \email{pavlidou@physics.uoc.gr}}

\author{J. Anton Zensus\\
         Max-Planck-Institut f\"ur Radioastronomie, Auf dem H\"ugel 69, Bonn 53121, Germany \\
        E-mail: \email{azensus@mpifr-bonn.mpg.de}}

      \abstract{The {\em RoboPol} program has been monitoring the $R$-band linear polarisation parameters of
        an unbiased sample of 60 gamma-ray-loud blazars and a ``control'' sample of 15 gamma-ray-quite
        ones. The prime drive for the program has been the systematic study of the temporal behaviour of the
        optical polarisation and particularly the potential association of smooth and long rotations of the
        polarisation angle with flaring activity at high energies. Here we present the program and discuss a
        list of selected topics from our studies of the first three observing seasons (2013--2015) both in the
        angle and in the amplitude domain.}

\FullConference{4th Annual Conference on High Energy Astrophysics in Southern Africa\\
		25-27 August, 2016\\
		Cape Town, South Africa}

\begin{document}

\def\aj{AJ}                   % Astronomical Journal
\def\actaa{Acta Astron.}      % Acta Astronomica
\def\araa{ARA\&A}             % Annual Review of Astron and Astrophys
\def\apj{ApJ}                 % Astrophysical Journal
\def\apjl{ApJ}                % Astrophysical Journal, Letters
\def\apjs{ApJS}               % Astrophysical Journal, Supplement
\def\ao{Appl.~Opt.}           % Applied Optics
\def\apss{Ap\&SS}             % Astrophysics and Space Science
\def\aap{A\&A}                % Astronomy and Astrophysics
\def\aapr{A\&A~Rev.}          % Astronomy and Astrophysics Reviews
\def\aaps{A\&AS}              % Astronomy and Astrophysics, Supplement
\def\azh{AZh}                 % Astronomicheskii Zhurnal
\def\baas{BAAS}               % Bulletin of the AAS
\def\bac{Bull. astr. Inst. Czechosl.}
                % Bulletin of the Astronomical Institutes of Czechoslovakia 
\def\caa{Chinese Astron. Astrophys.}
                % Chinese Astronomy and Astrophysics
\def\cjaa{Chinese J. Astron. Astrophys.}
                % Chinese Journal of Astronomy and Astrophysics
\def\icarus{Icarus}           % Icarus
\def\jcap{J. Cosmology Astropart. Phys.}
                % Journal of Cosmology and Astroparticle Physics
\def\jrasc{JRASC}             % Journal of the RAS of Canada
\def\memras{MmRAS}            % Memoirs of the RAS
\def\mnras{MNRAS}             % Monthly Notices of the RAS
\def\na{New A}                % New Astronomy
\def\nar{New A Rev.}          % New Astronomy Review
\def\pra{Phys.~Rev.~A}        % Physical Review A: General Physics
\def\prb{Phys.~Rev.~B}        % Physical Review B: Solid State
\def\prc{Phys.~Rev.~C}        % Physical Review C
\def\prd{Phys.~Rev.~D}        % Physical Review D
\def\pre{Phys.~Rev.~E}        % Physical Review E
\def\prl{Phys.~Rev.~Lett.}    % Physical Review Letters
\def\pasa{PASA}               % Publications of the Astron. Soc. of Australia
\def\pasp{PASP}               % Publications of the ASP
\def\pasj{PASJ}               % Publications of the ASJ
\def\rmxaa{Rev. Mexicana Astron. Astrofis.}%
                % Revista Mexicana de Astronomia y Astrofisica
\def\qjras{QJRAS}             % Quarterly Journal of the RAS
\def\skytel{S\&T}             % Sky and Telescope
\def\solphys{Sol.~Phys.}      % Solar Physics
\def\sovast{Soviet~Ast.}      % Soviet Astronomy
\def\ssr{Space~Sci.~Rev.}     % Space Science Reviews
\def\zap{ZAp}                 % Zeitschrift fuer Astrophysik
\def\nat{Nature}              % Nature
\def\iaucirc{IAU~Circ.}       % IAU Cirulars
\def\aplett{Astrophys.~Lett.} % Astrophysics Letters
\def\apspr{Astrophys.~Space~Phys.~Res.}
                % Astrophysics Space Physics Research
\def\bain{Bull.~Astron.~Inst.~Netherlands} 
                % Bulletin Astronomical Institute of the Netherlands
\def\fcp{Fund.~Cosmic~Phys.}  % Fundamental Cosmic Physics
\def\gca{Geochim.~Cosmochim.~Acta}   % Geochimica Cosmochimica Acta
\def\grl{Geophys.~Res.~Lett.} % Geophysics Research Letters
\def\jcp{J.~Chem.~Phys.}      % Journal of Chemical Physics
\def\jgr{J.~Geophys.~Res.}    % Journal of Geophysics Research
\def\jqsrt{J.~Quant.~Spec.~Radiat.~Transf.}
                % Journal of Quantitiative Spectroscopy and Radiative Transfer
\def\memsai{Mem.~Soc.~Astron.~Italiana}
                % Mem. Societa Astronomica Italiana
\def\nphysa{Nucl.~Phys.~A}   % Nuclear Physics A
\def\physrep{Phys.~Rep.}   % Physics Reports
\def\physscr{Phys.~Scr}   % Physica Scripta
\def\planss{Planet.~Space~Sci.}   % Planetary Space Science
\def\procspie{Proc.~SPIE}       % Proceedings of the SPIE

\section{Introduction}

The variability of both the angle and the degree of optical polarisation has been known since the early
polarisation studies of quasars \cite{1966ApJ...146..964K}. After the first detection of a long monotonous
change of the optical polarisation plane \cite{1988A&A...190L...8K} -- nowadays termed as ``rotation'' or
``swing'' -- the attention of the community was caught when a similar event was found in the blazar BL
Lacertae \cite{2008Natur.452..966M}. In that particular case the rotation appeared to be associated with
structural evolution of the radio jet as well as flaring activity at high energies. The rotation of the
optical polarisation plane was interpreted as the natural consequence of an emission element following a
helical path. The sequence of events observed at the source (structural evolution, optical-to-gamma rays
outburst, and delayed radio flaring activity) led to the understanding that the region where the acceleration
takes place is dominated by a helical magnetic field thus providing insight on the inner parts of the jet.

As rotations of the electric vector position angle (EVPA) were being detected, a broad spectrum of alternative
possibilities were proposed to interpret the observations. Early studies have been predicting that such events
could naturally occur as a result of "random walks" caused by the evolution of a turbulent magnetic field
\cite{Jones1985ApJ}. Shocks travelling in non-axisymmetric jets have also been discussed as a possible
mechanism \cite{1985ApJ...289..188K}; flares with significant degree of polarisation occurring in the
accretion disc \cite{1993Ap&SS.206...55S}; systems with two independent sources of polarised emission
\cite{1982ApJ...260..855B} and jet bending \cite{2010Natur.463..919A} are among the most commonly discussed
proposals.

The {\em RoboPol}\footnote{https://robopol.org} high cadence blazar polarisation monitoring program was
initiated with the aim to conduct a systematic study of these events and explore their full potential to probe
the physical processes at the radiating plasma. That involved the design, the funding, the construction and
the operation of a 4-channel optical polarimeter.

In the following we provide a brief yet concise description of the program and summarise some among the
most noteworthy findings in terms of the angle of polarisation plane as well as the polarisation
amplitude. Most of the material that follows is included in a series of papers
\cite{2014MNRAS.442.1693P,2014MNRAS.442.1706K,2015MNRAS.453.1669B,2016MNRAS.457.2252B,2016MNRAS.462.1775B,2016MNRAS.463.3365A,2016arXiv160808440H}.

\section{The {\em RoboPol} program}

The possible association of the EVPA rotations with the activity in the MeV -- GeV energy range
(e.g. \cite{2010Natur.463..919A,2013ApJ...768...40L}) has clearly been the drive for the initiation of the
{\em RoboPol} program. Should such an association be proven physical it could provide insight on the physics
of high energy activity. The considerable efforts put on understanding these events prior to {\em RoboPol}
were primarily event-driven, inevitably resulting in biased datasets unsuitable for statistically rigorous
studies. Instead, the guiding principles for {\em RoboPol} have been
\cite{2014MNRAS.442.1693P,2014MNRAS.442.1706K}:
\begin{enumerate}
\item To create a database of rotation events by systematically observing a sample of gamma-ray blazars as
  unbiased as possible.
\item To compile a similar dataset for a gamma-ray quiet ``control'' sample of blazars ideally identical in
  all properties to the ``main'' sample (except for the gamma-ray loudness).
\item To carry out this exercise with a cadence (a) adequately broad to cover all possible rates of rotation
  and (b) adaptive do make most efficient usage of the available observing time.
\item To have all the necessary control on the systematics in order to achieve the necessary degree of
  accuracy.
\end{enumerate} Such a dataset would then provide the base for studying the temporal polarimetric behaviour of
blazars; assessing the coincidence of rotation events with the activity at high energies; investigating the
physical processes producing those events; examining the dependence of the polarisation parameters on source
properties; conducting statistically reliable population studies.

To accomplish the aims set by this challenging scientific case three requirements became apparent from the
very beginning: large amount of telescope time, a sensitive polarimeter with high degree of systematics
control and a carefully selected unbiased sample.
 
\paragraph{Large amount of telescope time.} The need for continuous monitoring of relatively large samples in
combination with the rotation rates that had been known from the literature made it obvious that a good
fraction of a week was imperative. That was made possible with the dedication of four nights per week by the
Skinakas\footnote{\url{http://skinakas.physics.uoc.gr}} 1.3-m telescope in Crete
\cite{2007Ippa....2b..14P}. This observing time has been applicable for the first three years of the {\em
  RoboPol} operations (2013 -- 2015) and for the part of the year that the observatory is operated (typically
March or April -- November).

\begin{figure}[] 
\centering
\begin{tabular}{cc}
  \includegraphics[trim={0 0 0 100},clip, width=0.52\textwidth]{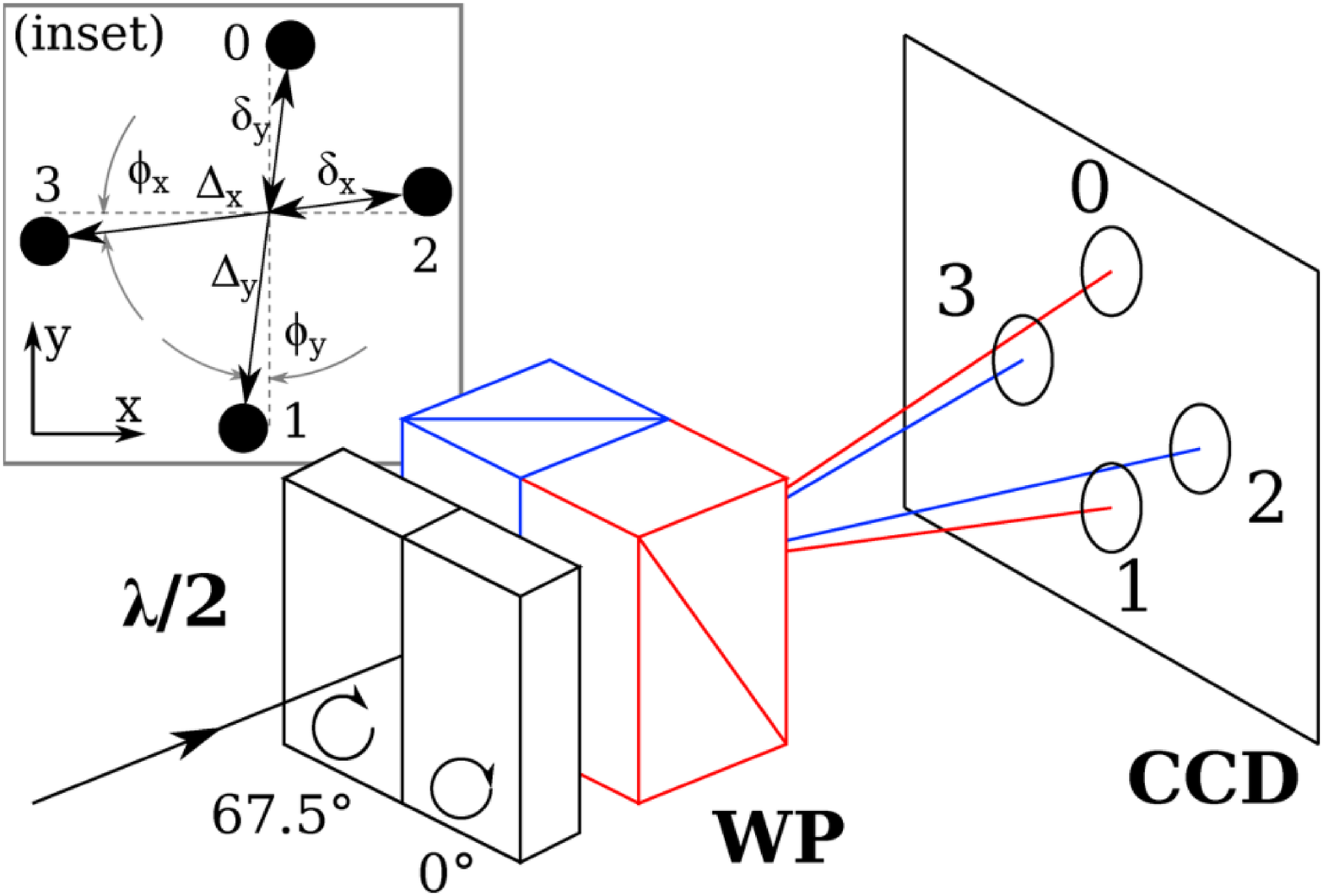} & \includegraphics[trim={0 0 0 100},clip, width=0.38\textwidth]{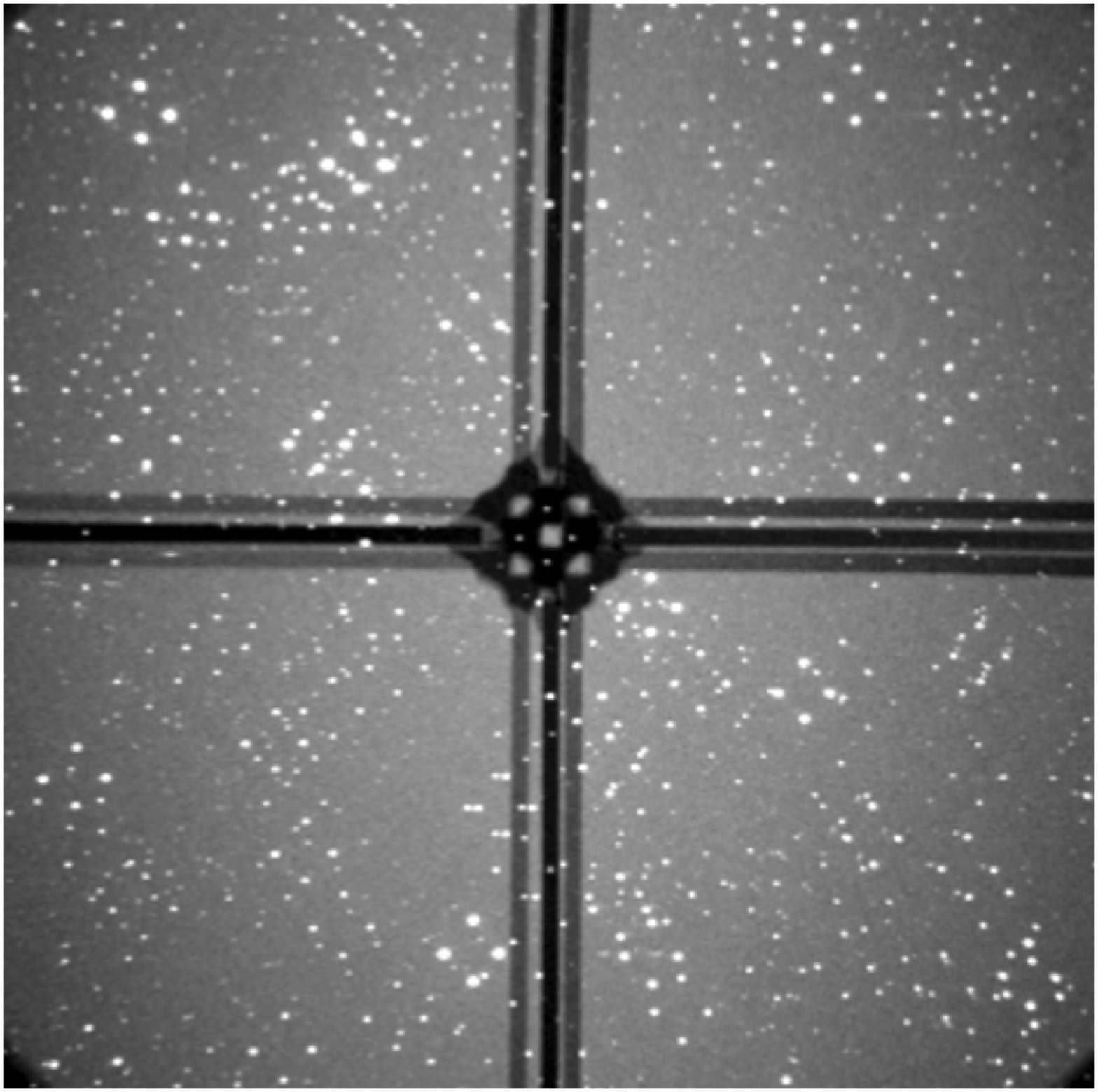} \\
\end{tabular}
\caption{{\it Left:} The design of the {\em RoboPol} polarimeter. An arrangement of half-wave retarders
  followed by two Wollaston prisms images each point source in the sky into four spots on the CCD plane. A
  simple operation on the counts from those four spots gives us the polarisation parameters. {\it Right:} An
  actual image of the entire field-of-view observed with {\em RoboPol}. The 1-to-4 mapping of point sources is
  obvious. The ``shadow'' of the central mask lowering the sky background in the vicinity of the central
  position is also obvious. Taken from \cite{2014MNRAS.442.1706K}.}
\label{fig:robopol}
\end{figure}

\paragraph{The {\em RoboPol} polarimeter.} A key element in the success of the {\em RoboPol} program has been
the four-channel {\em RoboPol} polarimeter (Fig.~\ref{fig:robopol}, Ramaprakash et al. in preparation). The
first essential ingredient in its design has been the complete absence of moving parts (apart from the filter
wheel which potentially can rotate). Contrary to many conventional polarimeters, {\em RoboPol} features a
combination of half-wave retarders followed by two Wollaston prisms in such an arrangement that each point
source in the sky is mapped into four spots on the CCD plane (see Fig.~\ref{fig:robopol}). Conducting
photometry of the four spots gives us directly the angle and degree of polarisation as it is described in
Section~2 of \cite{2014MNRAS.442.1706K}. This design minimises the systematics. The second ingredient that
improves the sensitivity of the instrument, allowing it to measure the polarisation vectors of weak sources,
is the central mask: a ``negative'' pattern that shadows the regions of the CCD where the four spots of the
central object are projected on, lowering the sky background by a factor of 4. Another essential part for the {\em
  RoboPol} smooth operation has been a sophisticated data acquisition and reduction ``pipeline'' which
conducts a series of automatic tasks, outputting the fractional Stokes parameters ($q=Q/I$ and $u=U/I$)
without any human intervention, as it is discussed by \cite{2014MNRAS.442.1706K}.
% -----------------------------------------------------------------------
% -----------------------------------------------------------------------

\paragraph{Unbiased sample.} As discussed earlier, the merit of the {\em RoboPol} findings
relies on the fact that the observed sample has been the result of non-biasing cuts applied on a gamma-ray
photon-flux-limited sample.

{\it The ``main'' sample; gamma-ray-loud (GL) sources:} Since the primary drive for the program has been the
potential association of the polarisation angle swings and the activity at high energies, the ``main'' sample
was drawn from the {\it Second Fermi Large Area Telescope Source Catalog} (2FGL,
\cite{2012ApJS..199...31N}). Starting from the 2FGL photon-flux-limited list we first applied a photon flux
cut ($F\left(E> 100 MeV\right) > 2\times10^{-8}$cm$^{-2}$s$^{-1}$) to avoid biases from the energy dependent
sensitivity of the detector, and excluded sources in the proximity of the galactic plane (galactic latitude
$|b|>10^{\circ}$) to minimise possible biases caused by imperfect knowledge of the galactic diffuse
background. From the resulting sub-set of 557 sources we selected those that (a) would rise to a maximum
elevation of more the $40^{\circ}$ at Skinakas for at least 90 consecutive nights in the window June --
November and (b) had an archival $R$-band magnitude of less than 17.5. This sub-set was finally examined for
field quality and observational constrains to create a list of 62 sources that comprised the ``main''
gamma-ray-loud sample that was monitored.

{\it The ``control'' sample; gamma-ray-quiet (GQ) sources:} The 15~GHz OVRO monitoring program showed that the
gamma-ray-loudness is related to the radio variability amplitude so that GQ sources show lower modulation
indices \cite{2011ApJS..194...29R}. In order to examine whether the optical polarisation properties are also
dependent on the gamma-ray-loudness, we selected the most radio-variable sources of the OVRO catalog that were
absent from the 2FGL. After applying similar cuts as for the ``main'' sample we compiled a list of 15
gamma-ray-quiet sources that comprised our ``control'' sample.

The two samples were observed as similarly as possible so that, despite their different sizes, a rigorous
comparison could be carried out. The details of the sample selection are discussed in
\cite{2014MNRAS.442.1693P}.

\subsection{Observations and dataset}
The {\em RoboPol} program was operated in its designed ``regular'' mode for the observing seasons of 2013,
2014 and 2015. The observing cadence was adopted to the source state of activity within a range from one
measurement every few days to a few measurements per night. The observations were done in the $R$-band.  The
observing cycle included a few-second-long exposures for (a) target acquisition optimisation and (b)
estimation of the exposure necessary for achieving the desired level of accuracy. The figure of merit was
reaching an SNR of 10 for a 3\% polarised source of $R= 17.5$~mag in no more than 30 minutes.

The overall uncertainty in the polarisation fraction is of the order of 0.01 and that in the polarisation
angle of the order of 1--10 degrees depending on the source brightness and polarisation. The photometry is
done using all standard stars found in any given field and the overall error in the $R$-band magnitude is
around 0.04 mag \cite{2014MNRAS.442.1693P,2015MNRAS.453.1669B,2016MNRAS.463.3365A}.

In 2016 the limited fraction of the telescope time that was granted for {\em RoboPol} observations was
dedicated to a list of 27 sources of the main sample that were most probable to show a rotation event. With
the reduced sample we significantly increased the cadence of observations for each monitored source, allowing
for an unprecedented time sampling of rotation events. This study will be discussed in a future publication.

\section{Polarisation angle rotations}
\begin{figure}[b] 
\centering
\begin{tabular}{cc}
   \includegraphics[trim={35 30 35 30},clip, width=0.49\textwidth]{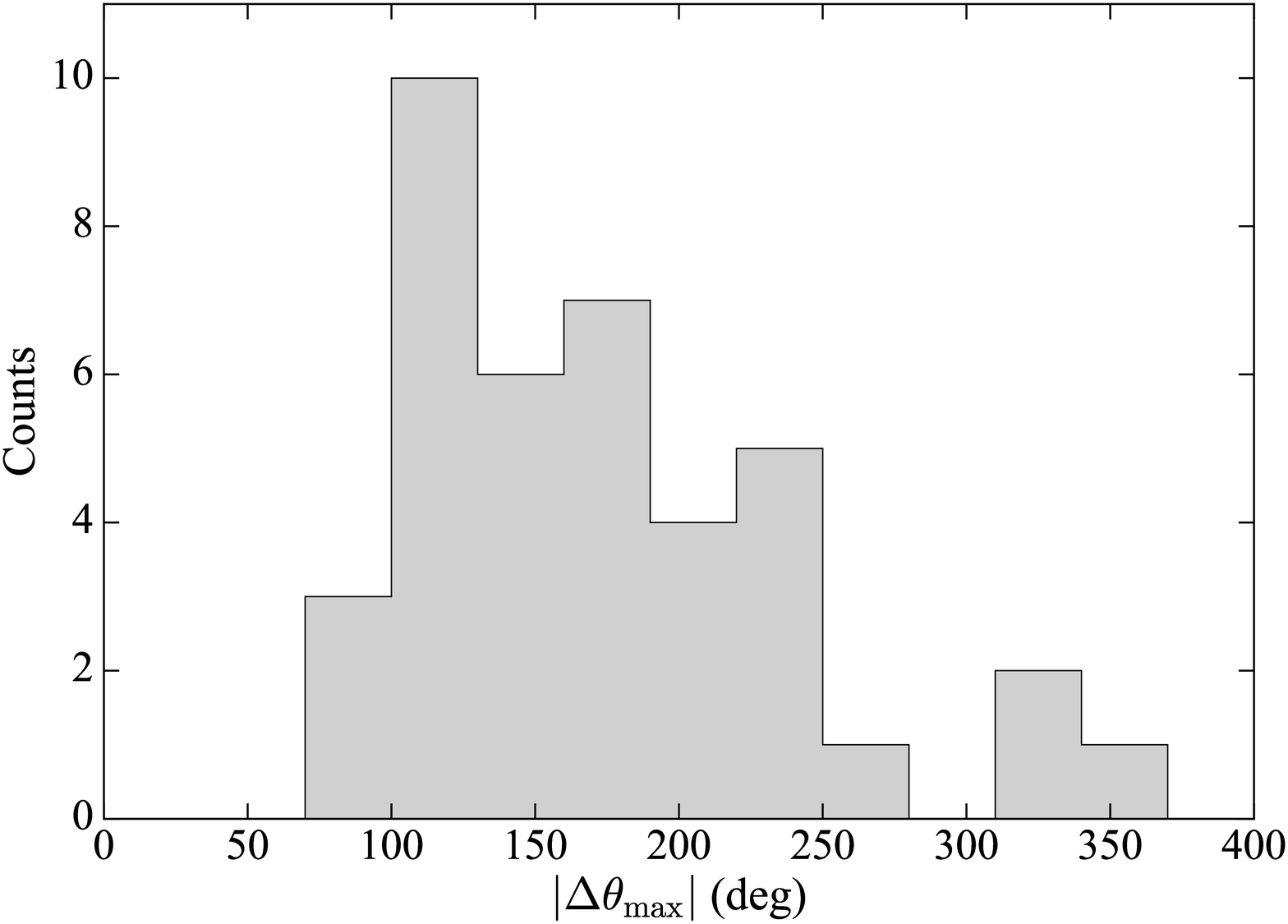}  &   \includegraphics[trim={35 30 30 30},clip,width=0.48\textwidth]{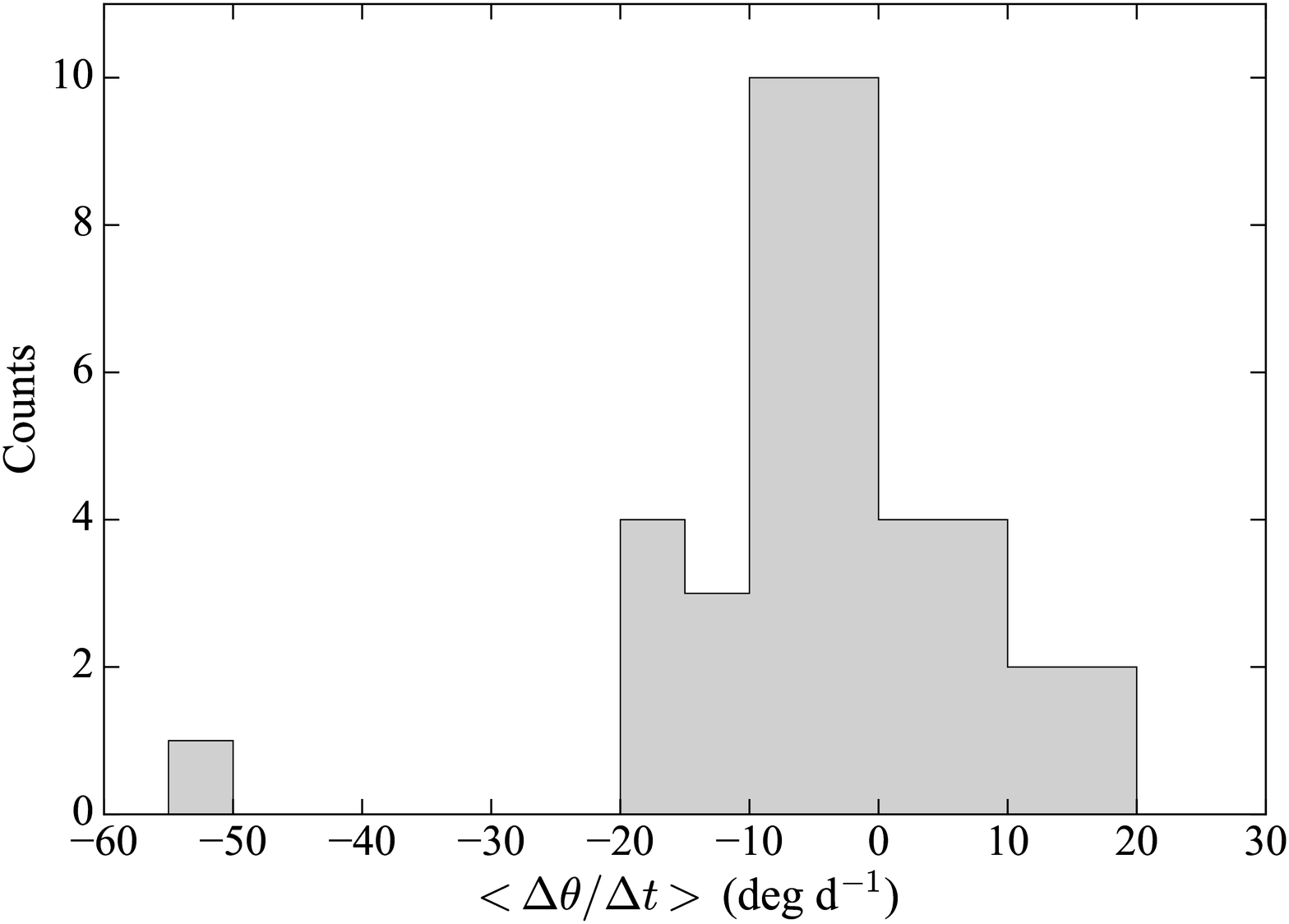}  \\
\end{tabular}
\caption{The distributions of two characteristic parameters of the observed rotations. {\it Left}, the
  rotation magnitude and {\it right} the average rotation rate. For the plots all 40 events in
  Table~\protect\ref{tab:rotations} have been included.}
\label{fig:rot_par}
\end{figure}

During the first three observing seasons (2013 -- 2015) {\em RoboPol} detected a total of 40 EVPA rotations in
24 blazars \cite{2015MNRAS.453.1669B,2016MNRAS.457.2252B,2016MNRAS.462.1775B} giving already a crude sense of
their frequency of occurrence. At least four measurements with significant angle swings between them are
required to define an EVPA rotation. The details are discussed in \cite{2015MNRAS.453.1669B}. The rotations
detected during the first (2013), second (2014) and third (2015) season are shown in Figures~2 of
\cite{2015MNRAS.453.1669B}, 3 of \cite{2016MNRAS.457.2252B} and 1 of \cite{2016MNRAS.462.1775B}. All 40 events
are listed in Table~\ref{tab:rotations} along with their most important parameters. All the data listed there
are taken from \cite{2015MNRAS.453.1669B,2016MNRAS.457.2252B,2016MNRAS.462.1775B}. In Figure~\ref{fig:rot_par}
we show the distributions of two characteristic parameters of the observed rotations: the rotation magnitude
and the average rotation rate.
% -----------------------------------------------------------------------
% -----------------------------------------------------------------------

%For the determination of an EVPA rotation first the $180^\circ$ ambiguity is resolved by assuming that the
%temporal variation is smooth and does not exceed the $90^\circ$ for two consecutive observation $n$ and
%$n+1$. The uncertainties in the angle estimation are accounted for by taking as significant angle variations
%that satisfy the condition $\Delta\theta=|\theta_{n+1}-\theta_{n}|>\sqrt{\sigma_{n+1}^2+\sigma_n^2}$ where
%$\sigma$ the corresponding angle uncertainty. If it happens to be
%$\Delta\theta - \sqrt{\sigma_{n+1}^2+\sigma_n^2}>90^\circ$ the angle $\theta_{n+1}$ is shifted by the integer
%multiples of $\pi$ that minimises $\Delta\theta$. 

\subsection{The first observing season (2013)}
The completion of the first observing season (2013, \cite{2015MNRAS.453.1669B}) revealed 16 rotations almost
doubling the sample of known events until then. Already that dataset led to a list of noteworthy
findings:

\paragraph{Rotations in GL and GQ sources.} Interestingly, the EVPA rotations were exclusively detected in
GL sources suggesting a physical connection between the regular temporal variability of optical polarisation
angle and the production of high energy emission. The identical treatment by the {\em RoboPol} observing
scheme of GL and GQ sources excluded the possibility that this dichotomy is caused by differences in the
sampling. 

\paragraph{Rotations in blazar classes.} Concerning different subclasses of blazars (LSP, ISP and HSP  or FSRQs
and BLL Lacs) they showed the same probability of showing EVPA rotations. Similarly, there is no preference
towards blazars known as TeV emitters or not.

\paragraph{Multiple rotations.} Eight blazars showed more that one rotation during this first observing
season. In these cases the rotations can show both senses of rotation, while the angle may rotate at a pace
very different (rotation rates may differ by a factor larger than two).

\paragraph{Rotations and gamma-ray activity.} The examination of the high energy activity during those 16
events showed that the gamma-ray photon fluxes on average do not show any systematic increase during the EVPA
rotations. However, in those cases where EVPA rotations were accompanied by a gamma-ray flare, the strongest
flares appear more contemporaneously with the corresponding rotation, whereas less intense flares show a
larger range of time lags between flare and rotation. This is shown in Figure~8 of
\cite{2015MNRAS.453.1669B}. If that result were to be confirmed with larger rotation datasets, it would
indicate the presence of two families of events: (a) rotation events associated with practically synchronous
and bright gamma-ray flares, and (b) rotations that are associated with much weaker gamma-ray outbursts that
can occur within a broad range of time lags from the rotation. Further, this would indicate two separate
mechanisms inducing EVPA rotations: one potentially deterministic process producing EVPA rotations and
higher-amplitude gamma-ray flares contemporaneously, and one potentially stochastic process producing
rotations which are not necessarily associated with the weaker gamma-ray flares occurring around the rotation
event.  This latter case could possibly be realised by random walk processes. Figure~8 of
\cite{2015MNRAS.453.1669B} will be revisited in an subsequent publication (Blinov et al. in preparation).

\paragraph{Random walks and EVPA rotations.} One of the mechanisms suggested to naturally produce smooth and
long rotations of the EVPA has been random walk processes in an evolving turbulent magnetic field
\cite{Jones1985ApJ}. We conducted Monte Carlo simulations to examine the capacity of this processes to
reproduce the phenomenologies we have observed and specifically the rotation magnitude
$\theta_\mathrm{max}$. We find that a random walk process can indeed produce rotations of the observed
magnitude in specific cases. It is however unlikely that the entirety of the observed rotations in the first
{\em RoboPol} season are due to a random walk process (probability $\le 1.5 \times 10^{-2}$,
\cite{2015MNRAS.453.1669B}).

\subsection{The second observing season (2014)}
During the second observing season (2014, \cite{2016MNRAS.457.2252B}) 11 new rotations were detected in 10
sources. The focus was put mostly on comparing the rotating and the non-rotating phases of the source as
well as estimating the rate of expected events. Some of the most noteworthy findings, were:
 
\paragraph{Optical flares and EVPA rotations.} Similarly to the question whether there is always a gamma-ray
flare accompanying an EVPA rotation, we have investigated the activity in the optical band during such
events. It appears that there is no association of rotations with contemporaneous optical flares. Figure~11 of
\cite{2016MNRAS.457.2252B} shows that the distribution of the ratio of the optical flux density during
rotation to that over the non-rotating period is peaking at unity; implying that there is no general increase
of the optical flux density during rotations. We also studied the flux density modulation index -- as a measure of the
variability amplitude -- during the rotation phase and outside of it to find that during the rotation the
variability properties remain -- on average -- unchanged as well.

\paragraph{Polarisation during EVPA rotations.} We further studied the behaviour of the degree of polarisation
during the rotations. Figure~9 in \cite{2016MNRAS.457.2252B} shows the ratio between the polarisation fraction
of 27 rotation events (from both the first and second seasons) during the rotation and that outside the
rotation. For more than 2/3 of the cases this ratio is below unity. The same exercise showed that the
modulation index of the polarisation fraction remains rather unchanged between the two states.

\begin{figure}[] 
\centering
  \includegraphics[trim={0 0 0 0},clip, width=1\textwidth]{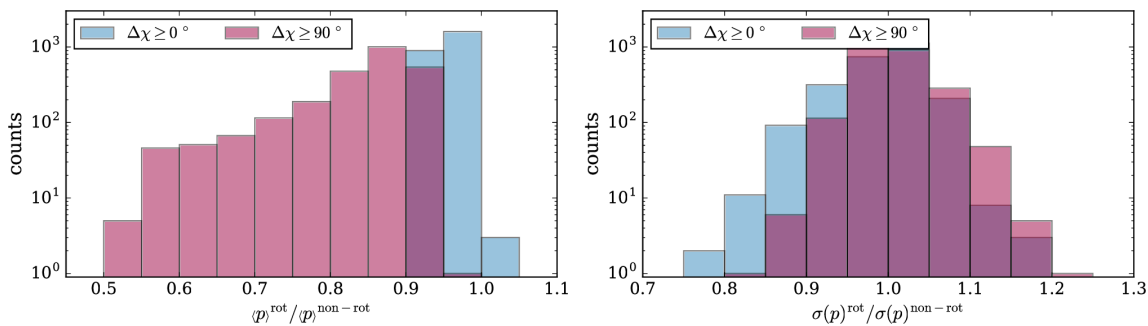} 
  \caption{ The prediction of the random walk model for the relative polarisation fraction {\it (left panel)}
    and the relative variability amplitude {\it (right panel)} during rotation and outside of it. The {\it
      blue} distributions correspond to rotations of arbitrarily large magnitude, while {\it red} ones
    correspond to rotations of more that $90^{\circ}$, only. The {\it red} distributions are directly
    comparable to Figure~9 in \cite{2016MNRAS.457.2252B}. The simulations represent idealised cases of very
    long continuous observations ($5\cdot10^5$ days, same time sampling as {\em RoboPol}, no seasonal
    gaps, Kiehlmann et al. in preparation). 
  }
\label{fig:rw_sim}
\end{figure}

We conducted large number of simulations to investigate the predictions of the random walk model
for these two quantities. Figure~\ref{fig:rw_sim} shows the relative polarisation fraction and the relative
variability amplitude during rotation and outside of it. As it is shown there the random walk model can
reproduced the observed behaviour even though the simulations refer to idealised cases of very long continuous
observations ($5\cdot10^5$ days, same time sampling as {\em RoboPol}, no seasonal gaps, Kiehlmann et al. in
preparation).
% -----------------------------------------------------------------------
% -----------------------------------------------------------------------

\begin{figure}[] 
\centering
  \includegraphics[trim={0 0 0 0},clip, width=0.6\textwidth]{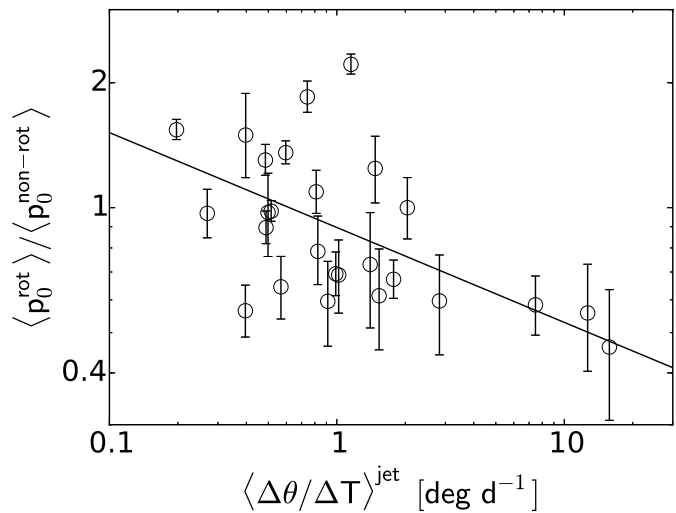} 
  \caption{The ``de-polarisation'' versus the emission-frame rotation rate. This is an update of Figure~10 in
    \cite{2016MNRAS.457.2252B} using all the available events detected within the first three regular
    seasons. The rotation rate has been corrected for Doppler blueshift and cosmological redshift.}
\label{fig:rel_pol_vs_rot_rate}
\end{figure}

\paragraph{Polarisation and the rotation rate.} Of great theoretical interest is the indication that the
relative polarisation fraction (during rotation and outside of it) anti-correlates with the rotation rates
measured at the emission frame i.e. corrected for Doppler blueshift and cosmological redshift (see Figure~10
in \cite{2016MNRAS.457.2252B}). Although the Doppler factors used there maybe be irrelevant to the optical
emission, their randomisation showed that a slope of the same significance can be produced in less than only
2\% of the cases. In Figure~\ref{fig:rel_pol_vs_rot_rate} we show the same plot updated to include all the
events detected with the first three seasons. The correlation coefficient is around $-0.535$ with a p-value of
around $4.9\cdot10^{-3}$ The fitted line has a slope of $-0.23\pm0.07$.
% -----------------------------------------------------------------------
% -----------------------------------------------------------------------

\subsection{The third observing season (2015)}
During 2015 another 13 rotation events were found in 10 blazars. On the basis of the entirety of events found
in all three regular seasons (2013 -- 2015) we conducted a thorough analysis presented in
\cite{2016MNRAS.462.1775B} which showed that the frequency of rotations varies significantly among
blazars. Additionally, no rotation was detected in a GQ blazar. Concerning the event rates we find that only
$\sim 28$\% of sources (collectively in GL and GQ samples) showed rotations with rates $\le
20$~deg~d$^{-1}$. The mean frequency of events is estimated to be one every 232~d (in the observer frame). The
rest of the sources (72\%) did not show rotations, constraining their frequency of occurrence (if any) to less than
one every 3230~d.
% -----------------------------------------------------------------------
\begin{table}[]
  \caption{The 40 rotations detected in the first three {\em RoboPol} observing seasons and some indicative
    parameters. Columns (1), (2): source identifiers; (3): source class:  LBL, IBL and HBL stand for
    high-frequency, intermediate-frequency and low-frequency peaked BL Lacs which FSRQ stand for flat spectrum
    radio quasar; (4): observing season duration; (5): EVPA
    rotation magnitude; (6) average rotation rate; (7) rotation duration; (8) observing season. The data
    are taken from \cite{2015MNRAS.453.1669B,2016MNRAS.457.2252B,2016MNRAS.462.1775B}.}
\small
  \label{tab:rotations}  
  \centering                    
  \begin{tabular}{lllrrrrr} 
    \hline\hline                 
    {\em RoboPol} ID  &Survey ID &Class  &\mc{1}{c}{$T_\mathrm{obs}$} &\mc{1}{c}{$\Delta\theta_\mathrm{max}$}
    &\mc{1}{c}{$\left<\frac{\Delta\theta}{\Delta t}\right>$} &\mc{1}{c}{$T_\mathrm{rot}$} &Season\\
                       &         & &\mc{1}{c}{(d)}             &\mc{1}{c}{(deg)}   &\mc{1}{c}{(deg/d)} &\mc{1}{c}{(d)} &   \\
    \hline
    \\
    RBPLJ0045$+$2127 &GB6 J0045$+$2127  &FSRQ     &113   &200    &4.2     &48  &2015  \\
    RBPLJ0136$+$4751 &OC 457            &FSRQ     &59    &$-$225 &$-$6.6  &34  &2013 \\
    \mc{1}{l}{\ldots}&\mc{1}{l}{\ldots} &\ldots   &136   &$-$92  &$-$2.2  &42  &2014 \\
    \mc{1}{l}{\ldots}&\mc{1}{l}{\ldots} &\ldots   &86    &$-$114 &$-$4.4  &26  &2015 \\
    \mc{1}{l}{\ldots}&\mc{1}{l}{\ldots} &\ldots   &\ldots      &$-$109 &$-$5.4  &20  &2015 \\
    RBPLJ1037$+$5711 &GB6 J1037$+$5711  &IBL      &54    &$-$165 &$-$5.3  &31  &2014  \\
    RBPLJ0259$+$0747 &PKS0256$+$075     &FSRQ     &72    &$-$180 &$-$4.8  &38  &2013 \\
    RBPLJ0721$+$7120 &S5 0716$$+$$71    &LBL      &88    &$-$208 &$-$14.8 &14  &2013 \\
    RBPLJ0854$+$2006 &OJ 287            &LBL      &51    &$-$154 &$-$6.7  &23  &2013 \\
    RBPLJ1048$+$7143 &S5 1044$$+$$71    &FSRQ     &142   &$-$188 &$-$9.0  &21  &2013  \\
    RBPLJ1512$-$0905 &PKS 1510$-$089    &FSRQ     &138   &243    &17.3    &14  &2014  \\
    \ldots           &\ldots            &\ldots   & \ldots     &$-$199 &$-$18.2 &11  &2014  \\
    \ldots           &\ldots            &\ldots   &115   &120    &6.0     &20  &2015 \\
    \ldots           &\ldots            &\ldots   & \ldots     &$-$97  &$-$7.0  &14  &2015 \\
    RBPLJ1555$+$1111 &PG 1553$$+$$113   &HBL      &129   &128    &5.6     &23  &2013 \\
    \ldots           &\ldots            &\ldots   &155   &145    &7.6     &19  &2014  \\
    RBPLJ1558$+$5625 &TXS1557$$+$$565   &IBL?     &137   &222    &7.2     &31  &2013 \\
    RBPLJ1635$+$3808 &4C 38.41          &FSRQ     &126   &$-$119 &$-$5.7  &21  &2015 \\
    RBPLJ1748$+$7005 &S4 1749$+$70      &IBL      &189   &$-$126 &$-$3.2  &39  &2014  \\
    RBPLJ1751$+$0939 &OT 081            &LBL      &177   &$-$335 &$-$10.5 &32  &2014  \\
    \ldots           &\ldots            &\ldots   &134   &$-$225 &$-$9.0  &25  &2015 \\
    RBPLJ1800$+$7828 &S5 1803$+$784     &LBL      &145   &$-$192 &$-$6.0  &32  &2014  \\
    \ldots           &\ldots            &\ldots   &147   &162    &2.9     &56  &2015 \\
    RBPLJ1806$+$6949 &3C 371            &LBL      &143      &238    &13.3    &18  &2013 \\
    \ldots           &\ldots            &\ldots   &143   &$-$347 &$-$16.5 &21  &2013 \\
    \ldots           &\ldots            &\ldots   &186   &$-$187 &$-$3.0  &63  &2014  \\
    RBPLJ1809$+$2041 &RX J1809.3$+$2041 &HBL      &130   &$-$427 &$-$4.7  &91  &2015 \\
    RBPLJ1836$+$3136 &RX J1836.2$+$3136 &FSRQ     &133   &182    &4.7     &39  &2015 \\
    RBPLJ1927$+$6117 &S4 1926$+$61      &LBL      &135   &$-$105 &$-$4.4  &24  &2013 \\
    RBPLJ2022$+$7611 &S5 2023$+$760     &IBL      &102   &107    &$-$4.7  &23  &2014  \\
    RBPLJ2202$+$4216 &BL Lac            &LBL      &137   &$-$253 &$-$51.0 &5   &2013 \\
    RBPLJ2232$+$1143 &CTA 102           &FSRQ     &140   &$-$312 &$-$15.6 &20  &2013 \\
    \ldots           &\ldots            &\ldots   &\ldots      &$-$140 &$-$11.8 &12  &2013 \\
    \ldots           &\ldots            &\ldots   &156   &$-$137 &$-$3.8  &36  &2015 \\
    RBPLJ2243$+$2021 &RGB J2243$+$203   &LBL      &169   &$-$183 &$-$5.9  &31  &2013 \\
    RBPLJ2253$+$1608 &3C 454.3          &FSRQ     &159   &$-$129 &$-$18.3 &7   &2013 \\
    \ldots           &\ldots            &\ldots   &156   &145    &16.3    &9   &2014  \\
    \ldots           &\ldots            &\ldots   &116   &$-$139 &$-$4.8  &29  &2015 \\
    \ldots           &\ldots            &\ldots   &\ldots      &101    &14.5    &7   &2015 \\
    RBPLJ2311$+$3425 &B2 2308$+$34      &FSRQ     &36    &74     &3.3     &23  &2013 \\\\
\hline                                                                         
  \end{tabular}                                                                
\end{table}                                       
% -----------------------------------------------------------------------

\section{The polarisation fraction}
\begin{figure}[] 
\centering
  \includegraphics[trim={10 0 0 0},clip, width=0.65\textwidth]{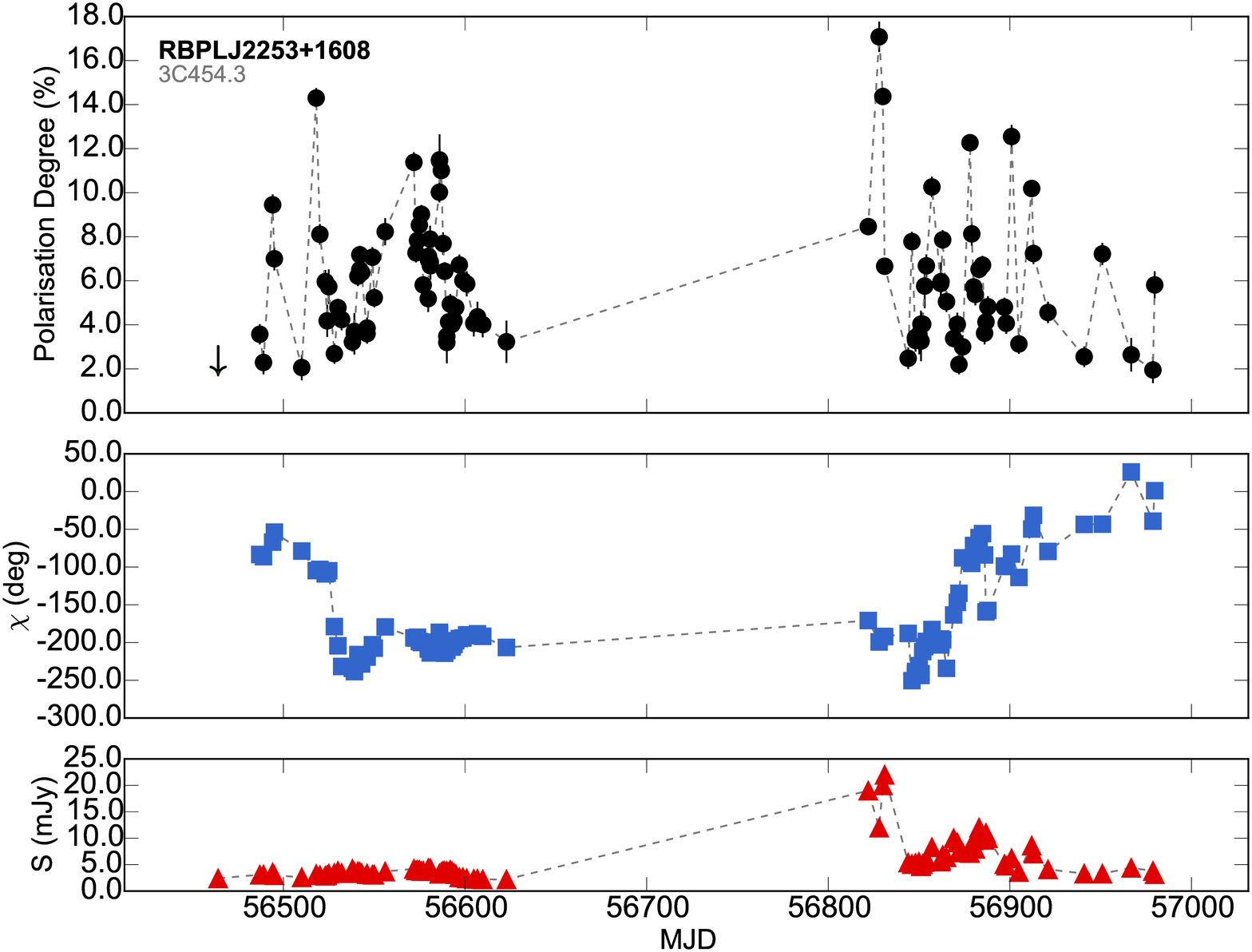} 
  \caption{An example of a {\em RoboPol} dataset for 3C454.3. From top to bottom: the polarisation degree (in
    percentage), the angle of polarisation (in degrees) and the $R$-band flux density (in mJy). This is the
    dataset from the first two seasons with a seasonal gap around MJD 56700.}
\label{fig:3c454}
\end{figure}

The {\em RoboPol} datasets include time series of the polarisation fraction (i.e., the degree of polarisation
normalised to unity), the polarisation angle and the $R$-band magnitude (or equivalently the flux density). An
example dataset if shown in Figure~\ref{fig:3c454}. 
% -----------------------------------------------------------------------
% -----------------------------------------------------------------------

In \cite{2016MNRAS.463.3365A} we focused on the degree of polarisation using the datasets collected during the
first two seasons. The main aim of our study was to understand the differences between the GL and GQ sources
and to investigate the factors that may be driving the occurrence of significant polarisation in blazars. Interestingly, our
findings led to an interpretation that could explain also some of the conclusions we reached in the studies of
the polarisation angle, as we discuss below.

\paragraph{The polarisation of GL and GQ sources.}
The single-shot measurements gathered during the {\em RoboPol} commissioning in early summer 2013 disclosed --
for the first time -- a significant separation of GL and GQ sources in terms of their polarisation degree
\cite{2014MNRAS.442.1693P}. We found is that that the population average of the polarisation fraction is more
than 6\% for the GL sources, but only slightly above 3\% for the GQ sources. Using the datasets of the first
two seasons that were potentially affected by source variability, updated these mean values to $9.2\pm0.8$\%
and $3.1\pm0.8$\%, respectively \cite{2016MNRAS.463.3365A}. For these latter estimates we have used maximum
likelihood polarisation fraction which accounts for statistical biases and observational uncertainties, under
the assumption that for both populations the polarisation fraction follows lognormal distributions.

\paragraph{The polarisation variability of GL and GQ sources.}
Unlike the degree of polarisation its variability amplitude is indistinguishable for GL and GQ sources as we
show in Figure~11 of \cite{2016MNRAS.463.3365A}. As we discuss there the two distributions cannot be
significantly different.

\paragraph{Polarisation and multi-band variability.}
The variability parameters and especially the variability amplitude could be seen as an indication of shocks
occurring in the emitting jet. Shocks imply compressions and a consequent local increase of the order of the
associated magnetic field. Thus, we expect an at least mild relation between the average degree of
polarisation and the variability in different bands as long as these assumptions hold. In Figures~8 and 9 of
\cite{2016MNRAS.463.3365A} we looked at the effect the radio and optical modulation index may have on the mean
polarisation fraction. Our study suggests that the median polarisation is higher, when the amplitude of
variability in radio or optical is larger. That is not however the case for the high energy (2FGL)
variability index. Finally, the variability amplitude of the polarisation itself seems to have no influence on
the median polarisation (Figure~10 therein).

\paragraph{Polarisation variability and redshift.} Despite the complete absence of a relation between the
median polarisation and redshift, we find a relation between the modulation index of the polarisation
fraction and redshift. When GL and the GQ sources are examined collectively a Spearman's test gave a $\rho$
of 0.43 with a p-value of $10^{-3}$. The positive correlation becomes further significant when upper limits of
the modulation index are used (see Figure~12 in \cite{2016MNRAS.463.3365A}). This implies a cosmic evolution
of the factors that contribute eruptive polarised emission.

\paragraph{Polarisation and synchrotron peak frequency.}
The key discovery of the study presented in \cite{2016MNRAS.463.3365A} is the dependence of the polarisation
fraction on the location of the SED synchrotron component peak which is shown in Figure~5 therein.  It becomes
evident that a synchrotron-peak-dependent envelope is binding the polarisation fraction so that
low-synchrotron-peaked (LSP) sources are on average more polarised than high-synchrotron-peaked (HSP) ones. At the same
time the polarisation spread decreases towards higher frequencies. As we discuss later this realisation led to
sketching a model that can explain a number of observable effects.
 
\paragraph{Randomness of the angle and synchrotron peak frequency.}
Furthermore, we observed that the randomness of the EVPA also depends on the location of synchrotron component
peak. LSP sources tend to show a random orientation of their EVPA, unlike HSP sources which tend to show a
preferred direction as it is demonstrated in Figure~7 of \cite{2016MNRAS.463.3365A}.

\section{A qualitative model for the interpretation of the observed trends}
\begin{figure}[] 
\centering
  \includegraphics[trim={0 15 0 0},clip, width=0.55\textwidth]{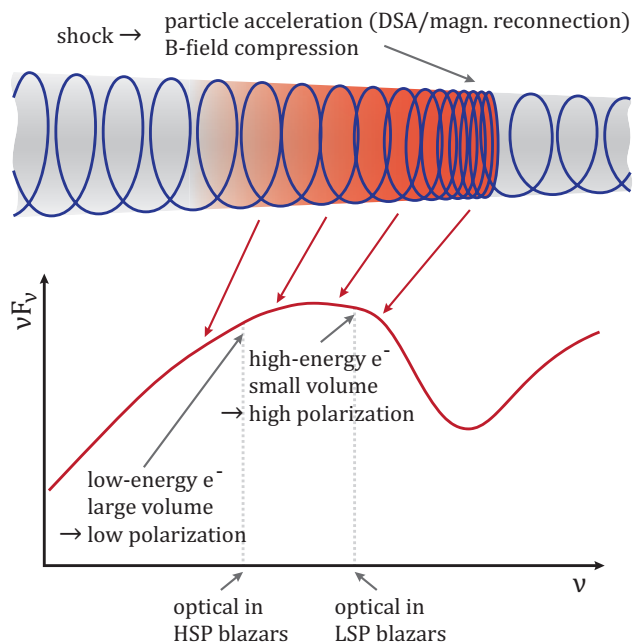} 
  \caption{A schematic representation of the qualitative model that can reproduce some of the important
    observations of the spectral energy distribution. The downstream direction is to the left.}
\label{fig:model}
\end{figure}

In Figure~\ref{fig:model} we show a simple qualitative model that seems able to explain most of the effects we
observe. It relies on the shock-in-jet scenario. The jet is assumed to be pervaded by a magnetic field in a
helical configuration, on which a turbulent component may be superposed. Mildly relativistic shocks may induce
efficient particle acceleration (e.g., via diffusive shock acceleration or magnetic reconnection) in a small
volume in the immediate downstream vicinity of the shock. Particles that are advected away from the shock on
their way downstream cool via synchrotron and Compton radiative losses. The most energetic particles that
produce the emission near and beyond the synchrotron peak are expected to be concentrated in a small volume
immediately downstream of the shock. There the shock-compressed magnetic field has a dominant ordered
(helical) component as well as shock-generated turbulent magnetic field. As a result, near and beyond the
synchrotron peak we expect substantial degrees of polarisation. Owing to progressive cooling of
shock-accelerated electrons as they are advected downstream, the volume from which lower frequency synchrotron
emission is received is expected to increase monotonically with decreasing frequency. The superposition of
radiation from zones with different B-field orientations causes de-polarisation so that one expects a lower
degree of polarisation with decreasing frequency.
 % -----------------------------------------------------------------------
% -----------------------------------------------------------------------

In the framework of this model the dichotomy between GL and GQ sources can be explained naturally. GL sources
are characterised by high degree of variability, indicating a jet dominance for most parts of the SED
due to a high degree of Doppler boosting and frequent occurrence of impulsive episodes of particle
acceleration as the ones described above. GQ sources on the other hand must be characterised by less extreme
Doppler boosting and less efficient particle acceleration events. This leads to energies lower than the ones
required for gamma-ray production at detectable levels. As a consequence, optical synchrotron emission is
likely to be produced on larger volumes than in the more active GL objects, naturally explaining the lower
degree of polarisation.

Even more direct is the explanation that the same model seems to be providing for the dependence of the degree
of polarisation on the synchrotron peak. In LSP blazars the synchrotron peak is typically located in the
infrared implying that $R$-band emission belongs to the high frequency part of the synchrotron emission. As we
discussed above, this is associated with high degree of polarisation. On the other hand, for HSP blazars the
synchrotron peak is located in UV or X-rays. In this case the optical emission is part of the low-frequency
synchrotron emission which is expected to give lower degrees of polarisation.

Within this framework, for LSP sources the region from which the optical emission emanates is characterised by
a dominant helical field component. In those cases then we expect that the rotations will occur (a)
deterministically and (b) practically concurrently with the gamma-ray flares.

\begin{figure}[] 
\centering
  \includegraphics[trim={0 0 0 0},clip, width=0.6\textwidth]{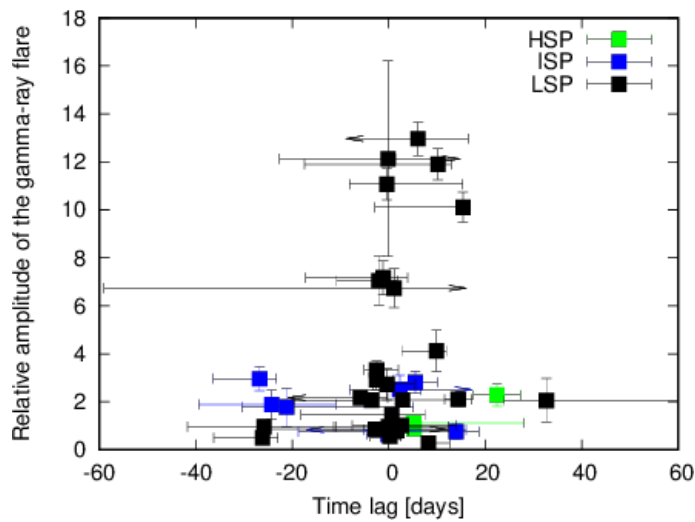} 
  \caption{The contemporaneity of a rotation to the corresponding gamma-ray flare against the gamma-ray flare
    amplitude. The $y$-axis shows the gamma-ray flare amplitude in arbitrary units and the $x$-axis the signed
  lag between the gamma-ray flare and the middle of the EVPA rotation.}
\label{fig:fig8}
\end{figure}

Some hint that this prediction may indeed be confirmed is given by Figure~\ref{fig:fig8}. There we duplicate
the plot shown in Figure~8 of \cite{2015MNRAS.453.1669B} except that here we use all available datasets. The
sources are labeled according to their synchrotron peak as LSP, ISP and HSP. All the events that are
synchronised with the brightest gamma-ray events have occurred exclusively in LSP sources. That is not the
case for the second population of events where there is an apparent mixing of source classes.
% -----------------------------------------------------------------------
% -----------------------------------------------------------------------

The combination of a helical and a turbulent component in the context of our assumed model seems to be giving
an explanation also for the observation that HSP sources tend to have preferred directions of the polarisation
angle while LSP sources incline to a random orientation.  The presence of a helical component means that both
LSP and HSP sources have an underlying stable EVPA component associated with it. In HSP sources where the
variability amplitudes are lower and variability time scales longer, the stable component may be more
dominant. That would explain why their EVPA is practically stable. It is essential however to note that only
long-term observations can confirm whether the EVPA has a truly preferred orientation on time scales longer
than the {\em RoboPol} observing periods.

In the case of HSP sources with preferred orientation of the polarisation angle it is only natural to expect
that this angle should align either perpendicular or parallel to the jet axis. We have confirmed that this is
indeed the case by studying the behaviour of TeV sources as it is demonstrated in Figure~7 of
\cite{2016arXiv160808440H}.

\section{The polarisation of high-energy BL Lac objects (TeV)}
In an attempt to understand what triggers the high energy emission in high energy BL Lac objects (TeV) we studied
a sample of 32 TeV and 19 non-TeV blazars within the framework of the {\em RoboPol} program. The study is in
press and currently available at \cite{2016arXiv160808440H} . 

With regards to the mean polarisation fraction the TeV-detected sources gave a value of around 5\% and non-TeV
sources around 7\%. This difference however is attributed to host galaxy contribution and is eliminated when
the host starlight is accounted for.  However, it appears that the TeV sources are characterised by slightly lower
fractional polarisation variability amplitude as compared to the non-TeV ones. In fact there seem to be no
intrinsic differences in the polarisation properties of the TeV and non-TeV sources. 

What is very interesting is that for TeV sources the EVPA tends to concentrate around preferred directions
while for non-TeV sources the angle tends to get rather random values. And for the majority of the cases the
EVPA and jet are aligned to less the $20^{\circ}$ implying a B-field perpendicular to the jet direction (see
Figure~7 in \cite{2016arXiv160808440H}). This has been seen as an important confirmation of our qualitative model.

\section{Conclusions}

Here we have attempted a brief summary of the most noteworthy findings from the first three observing seasons
of the {\em RoboPol} program. Between 2013 and 2015 {\em RoboPol} has created the first unbiased set of EVPA
rotations that provided the ground for the first systematic study of these events and gain understanding on
the mechanisms that produce them. Already from the first season it became clear that all classes of blazars
can rotate. With the dataset from the first two seasons we confirmed that GL are more polarised -- as a
population -- than GQ sources. The polarisation as well as its spread depend on the synchrotron peak while the
EVPA clearly shows a preferred direction for HSP sources. A jet populated by a helical field superposed to a
turbulent component and with impulsive events of particle acceleration can provide a natural explanation for
most of the observed phenomenologies.

\section*{}

The {\em RoboPol} program is a collaboration between: {\it University of Crete (Heraklion, Greece)}, {\it
  Max-Planck-Institut f\"ur Radioastronomie (Germany)}, {\it California Institute of Technology (USA)}, {\it
  Inter-University Centre for Astronomy and Astrophysics (India)} and the {\it Torun Centre for Astronomy,
  Nicolaus Copernicus University (Poland) }.

%\begin{thebibliography}{99}
%\bibitem{...}
%....

%\end{thebibliography}
%\bibliographystyle{JHEP} % style aa.bst
%\bibliography{/Users/mangel/work/Literature/MyBIB/References.bib} % your references Yourfile.bib

\providecommand{\href}[2]{#2}\begingroup\raggedright\endgroup

\end{document}